\documentclass[aps,pre,twocolumn]{revtex4-2}

\usepackage[dvipdfmx]{graphicx}
\usepackage{amsmath}
\usepackage{bm}
\usepackage{color}

\begin{document}

\title{Lateral transport of domains in anionic lipid bilayer membranes under DC electric fields: A coarse-grained molecular dynamics study}

\author{Hiroaki Ito}
\email[]{ito@chiba-u.jp}
\affiliation{Department of Physics, Graduate School of Science, Chiba University, Chiba 263-8522, Japan}

\author{Naofumi Shimokawa}
\affiliation{School of Materials Science, Japan Advanced Institute of Science and Technology, Ishikawa 923-1292, Japan}

\author{Yuji Higuchi}
\affiliation{Research Institute for Information Technology, Kyushu University, Fukuoka 819-0395, Japan}

\begin{abstract}
    Dynamic lateral transport of lipids, proteins, and self-assembled structures in biomembranes plays crucial roles in diverse cellular processes. In this study, we perform a coarse-grained molecular dynamics simulation on a vesicle composed of a binary mixture of neutral and anionic lipids to investigate the lateral transport of individual lipid molecules and the self-assembled lipid domains upon an applied direct current (DC) electric field. Under the potential force of the electric field, a phase-separated domain rich in the anionic lipids is trapped in the opposite direction of the electric field. The subsequent reversal of the electric field induces the unidirectional domain motion. During the domain motion, the domain size remains constant, but a considerable amount of the anionic lipids is exchanged between the anionic-lipid-rich domain and the surrounding bulk. While the speed of the domain motion (collective lipid motion) shows a significant positive correlation with the electric field strength, the exchange of anionic lipids between the domain and bulk (individual lipid motion) exhibits no clear correlation with the field strength. The mean velocity field of the lipids surrounding the domain displays a two-dimensional (2D) source dipole. We revealed that the balance between the potential force of the applied electric field and the quasi-2D hydrodynamic frictional force well explains the dependence of the domain motions on the electric-field strengths. The present results provide insight into the hierarchical dynamic responses of self-assembled lipid domains to the applied electric field and contribute to controlling the lateral transportation of lipids and membrane inclusions.
\end{abstract}

\maketitle

\section{INTRODUCTION}

Dynamic lateral arrangement of lipid molecules and membrane proteins in cell membranes plays crucial roles in various cellular processes such as signal transduction, membrane trafficking, and energy conversion. The cellular processes associated with cell membranes are believed to be facilitated by forming small functional domains called lipid rafts, in which specific lipid molecules and membrane proteins are dynamically self-assembled\cite{Lingwood2010,Levental2020}. The transiently formed small rafts can further coalesce into a large cluster\cite{Simons2011}. In this process, small domains are laterally transported in the bilayer membrane to contact each other at a distance short enough to attract by lipid-lipid and protein-protein interactions. As a pioneering demonstration in plasma membranes, clustering of the raft ganglioside GM1 by the cross-linking mediated by cholera toxin has been observed at a physiological temperature\cite{Lingwood2008}. For a deeper understanding of physical principles underlying the membrane-associated transport phenomena and potential applications of these cellular processes, it is necessary to investigate the mechanism of the lateral transport of individual molecules as well as the transport of the self-assembled lipid domains themselves.

Artificial lipid bilayer systems such as giant unilamellar vesicles (GUVs) and supported lipid bilayer (SLB) membranes are suitable platforms for studying the fundamental mechanism of domain formation \cite{Stottrup2004,Veatch2002}. The lateral transport phenomena of lipids\cite{Lindblom2009,Machan2010} and raft-like macroscopic domains\cite{Saeki2006,Yanagisawa2007,Cicuta2007,Stanich2013} in lipid bilayer membranes have been studied using such systems, and the influences of lipid species, aqueous solutions, coupling between the membrane and solutions, etc., on the lateral diffusivity have been discussed. As one of the fundamental attempts to quantify the lateral transport by the mobility of the membrane inclusions, electrical manipulation of charged lipids and proteins in lipid bilayer membranes by applying a tangential direct current (DC) electric field has been developed. The technique was first adopted to observe the redistribution of a charged complex, a membrane receptor with a charged ligand concanavalin A, in the cell membrane\cite{Poo1977,McLaughlin1981} and recently to demonstrate the lateral migration of lipid rafts and orienting the cell migration\cite{Lin2017}. The underlying mechanism of the lateral motion of membrane inclusions was then extensively studied using the SLB membranes\cite{Stelzle1992,Groves1995}. In most situations of lipid bilayer membranes floating in a three-dimensional (3D) solvent with counterions and salts, the charged membrane inclusions are driven by electrophoresis and electroosmosis\cite{Stelzle1992}. These two effects are typically competing, and the dominant factor depends on the geometry of the inclusions and the concentration of electrolytes in the solvent. The two-dimensional (2D) movement of the charged membrane inclusions can be well-described by the advection-diffusion equation and results in the concentration gradient with an exponential profile in a steady state\cite{Groves1995}, indicating that the individual inclusions are independently dragged by the electric field under the thermal fluctuation. The resultant redistribution of the charged species in the SLB membranes has been utilized for the measurement of diffusivity or charge of the inclusions\cite{Stelzle1992,Groves1995,Huang2022} and separation of membrane proteins\cite{Liu2011}.

Although the mechanism and applications of the electric-field-induced concentration gradient of charged membrane inclusions have been intensively studied, the response of the self-assembled domains to the electric field is not fully understood due to the complexity of the self-assembled hierarchical structure with long-range electrostatic interaction. During the past decade, phase separation of anionic phospholipids in GUVs has attracted increasing attention\cite{Shimokawa2010,Blosser2013,Pataraia2014,Guo2021} because the cell membranes\cite{Yeung2008} and organelle membranes of lysosomes, mitochondria, etc., contain anionic phospholipids. In this context, a manipulation technique of the phase-separated charged domains on a GUV by applying an external DC electric field has been proposed\cite{Zendejas2011}. In this demonstrative experiment, the electrophoresis dominated the domain dynamics; the charged domains are dragged by a DC electric field and oriented to the direction of the electric field within seconds. Toward a deeper understanding of charge-regulated hierarchical structure formation and applications of emergent functions, it is necessary to elucidate multiscale dynamics ranging from microscopic molecules to mesoscopic domains and the dependence of the mobility on the strengths of electric fields.

For revealing the dynamics of lipid molecules and self-assembled domains at the molecular level, molecular simulations are helpful\cite{Noguchi2009,Wassenaar2015,Marrink2019,Higuchi2023}. In particular, the highly coarse-grained model allows us to reproduce macroscopic phase separation for neutral lipid vesicles\cite{Cooke2005} and anionic lipid vesicles with the interaction potential based on the Debye-H\"{u}ckel theory\cite{Himeno2015,Ito2016,Shimokawa2019}. In this paper, we report domain responses to external electric fields by coarse-grained molecular dynamics (MD) simulation. The details of the simulation are explained in Sec.~\ref{methods}. In the present model, we considered the electrostatic repulsive interaction between anionic head groups of lipids with the Debye-H\"{u}ckel approximation and the potential force of homogeneous DC electric fields. The results of vesicle dynamics are shown in Sec.~\ref{results}. We observed domain formation under the applied electric  and checked the domain response upon the reversal of the direction of the electric fields with various strengths. From statistics of lipid species and analysis of the mean velocity field around the domain, we analyzed the microscopic and mesoscopic dynamics of each lipid species during the domain motion. In Sec.~\ref{discussion}, the domain motion and its field-strength dependence are discussed as hydrodynamic drag problems for a quantitative understanding of orienting the charged domain under an externally applied electric field.

\section{METHODS}
\label{methods}
In our coarse-grained MD simulation, a single lipid molecule is represented by three beads: linearly connected one hydrophilic bead and two hydrophobic beads, which correspond to the lipid head group and hydrocarbon chains, respectively. The excluded volume interaction between two beads separated by a distance $r$ is
\begin{equation}
    V_\mathrm{ex}(r) = 
    \left\{
        \begin{array}{ll}
            4v
            \left[
                \left(\frac{b}{r}\right)^{12}-\left(\frac{b}{r}\right)^6+\frac{1}{4}
            \right],
            &
            r\leq r_\mathrm{c},
            \\
            0,
            &
            r>r_\mathrm{c},
        \end{array}
    \right.
    \label{eq:V_ex}
\end{equation}
where $r_\mathrm{c} = 2^{1/6}b$. $v = k_\mathrm{B}T$ is the unit of energy, where $k_\mathrm{B}$ and $T$ are the Boltzmann constant and absolute temperature, respectively. For the bilayer stability, we chose the parameter $b$ for three combinations of the bead species as $b_\mathrm{head,head} = b_\mathrm{head,tail} = 0.95\sigma$ and $b_\mathrm{tail,tail} = \sigma$, where $\sigma = 7.09~\mathrm{\AA}$ is the typical cross-sectional diameter of a single lipid molecule as the unit of length. The potentials for the stretching and bending of a bond between two connected beads are
\begin{equation}
    V_\mathrm{stretch}(r) = \frac{1}{2}k_\mathrm{stretch}(r-\sigma)^2
    \label{eq:V_stretch}
\end{equation}
and
\begin{equation}
    V_\mathrm{bend}(\theta) = \frac{1}{2}k_\mathrm{bend}(1-\cos\theta)^2,
    \label{eq:V_bend}
\end{equation}
where $k_\mathrm{stretch} = 500v/\sigma^2$ and $k_\mathrm{bend} = 60v$ are the bonding strength of the connected beads and the bending stiffness of a lipid molecule, respectively. Here, $0\leq\theta\leq\pi$ is the angle between two adjacent bonds. The attractive hydrophobic interaction between hydrophobic beads is
\begin{equation}
    V_\mathrm{attr}(r) = 
    \left\{
        \begin{array}{ll}
            -v,
            &
            r<r_\mathrm{c},
            \\
            -v\cos^2\left[\frac{\pi(r-r_\mathrm{c})}{2w_\mathrm{c}}\right],
            &
            r_\mathrm{c}\leq r\leq r_\mathrm{c}+w_\mathrm{c},
            \\
            0,
            &
            r>r_\mathrm{c}+w_\mathrm{c},
        \end{array}
    \right.
    \label{eq:V_attr}
\end{equation}
where $w_\mathrm{c}$ is the phenomenological cutoff length of the attractive interaction. The lipid membranes are in ``gel'' or ``liquid'' phases depending on $w_\mathrm{c}$. In this study, we adopted $w_\mathrm{c}/\sigma = 1.7$ for the neutral-neutral pairs and the anionic-anionic pairs and $w_\mathrm{c}/\sigma = 1.5$ for the neutral-anionic pairs to induce the phase separation with the binary lipid mixture. Note that $w_\mathrm{c}/\sigma = 1.7$ and $1.5$ respectively correspond to ``gel'' and ``liquid'' phases, and the ``gel'' phase with $w_\mathrm{c}=1.7$ is near the boundary of the ``gel'' and ``liquid'' phases in the parameter space\cite{Cooke2005,Ito2016}. Thus, the lipid molecules in the membrane are still mobile, and the phase-separated domains become circular by the interfacial energy, which is characteristic of a so-called liquid-disordered phase.
To represent the anionic lipids, we considered the electrostatic repulsive interaction between anionic head groups. The repulsive electrostatic interaction is described as the Debye-H\"{u}ckel potential
\begin{equation}
    V_\mathrm{rep}(r) = v\ell_\mathrm{B}q_1q_2\frac{\exp(-r/\ell_\mathrm{D})}{r},
    \label{eq:V_rep}
\end{equation}
where $\ell_\mathrm{B} = \sigma$ is the Bjerrum length, $q_1$ and $q_2$ are the valencies of the interacting charged head groups, and $\ell_\mathrm{D} = \sigma\sqrt{\epsilon k_\mathrm{B}T/n_0e^2}$ is the Debye screening length. $n_0$, $\epsilon$, and $e$ are the bulk salt concentration, the dielectric constant of the solution, and the elementary charge, respectively. We set $n_0=100\,\mathrm{mM}$ and $q_1=q_2=-1$, which represent a typical condition for monovalent anionic lipids in an aqueous solution of a physiological monovalent salt concentration. We did not set any cutoff for the screened electrostatic repulsion.
We imposed the static direct current (DC) electric field $\mathbf{E}=Ev\mathbf{e}_z$ along the $z$-axis in the Cartesian coordinates, where $E$ and $\mathbf{e}_z$ represent the strength of the DC electric field and the unit vector in the $z$-direction, respectively. The potential of the DC electric field is, therefore,
\begin{equation}
    V_\mathrm{EF}(r) = -q_iEvz.
    \label{eq:V_EF}
\end{equation}
The position of the $i$-th bead $\mathbf{r}_i$ obeys the stochastic dynamics described by the Langevin equation
\begin{equation}
    m\frac{\mathrm{d}^2\mathbf{r}_i}{\mathrm{d}t^2} = -\eta\frac{\mathrm{d}\mathbf{r}_i}{\mathrm{d}t} + \mathbf{f}_i^V + \bm{\xi}_i,
\end{equation}
where $m=1$ and $\eta=1$ are the mass and the drag coefficient, respectively. The total potential force $\mathbf{f}_i^V$ is calculated from the sum of the derivatives of the interaction potentials described in Eqs. (\ref{eq:V_ex})--(\ref{eq:V_EF}). The Brownian force $\bm{\xi}_i$ satisfies the fluctuation-dissipation theorem
\begin{equation}
    \left<\xi_{i\alpha}(t)\xi_{j\beta}(t^\prime)\right> = 6v\eta\delta_{ij}\delta_{\alpha\beta}\delta(t-t^\prime),
    \label{eq:fdt}
\end{equation}
where $\delta_{ij}$ is the Kronecker delta, $\delta(\cdot)$ is the Dirac delta, and Greek indices denote the spatial coordinates. The time increment for solving the discretized equation is set at $\mathrm{d}t=7.5\times10^{-3}\tau$, where $\tau=\eta\sigma^2/v$ is the unit of time.

\begin{figure}[b]
    \centering
    \includegraphics{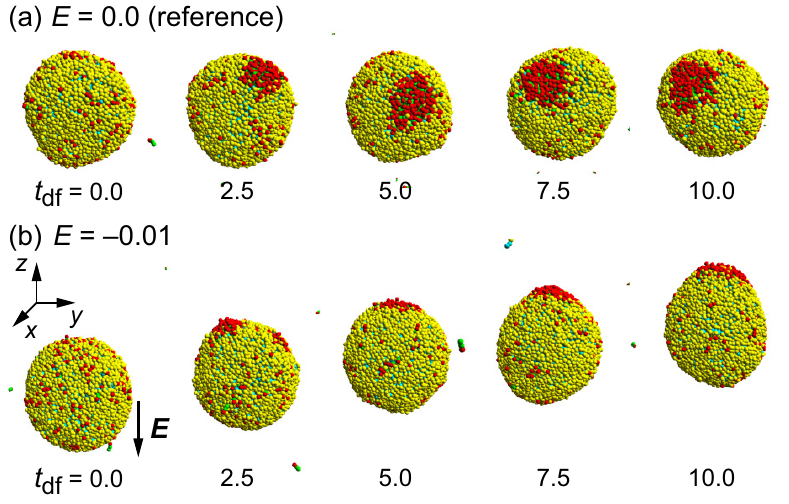}
    \caption{\label{fig1}(a) Typical snapshots of domain formation of anionic lipids (red) for $E = 0.0$. Calculation time for the domain formation $t_\mathrm{df}$ is represented with the unit of $7500\tau$. (b) Typical snapshots of domain formation of anionic lipids for $E = -0.01$.}
\end{figure}
    
In this study, we calculated the dynamics of bilayer vesicles composed of a binary mixture of anionic lipids and electrically neutral lipids. The spherical bilayer vesicle consists of 500 anionic lipids and 4500 neutral lipids, thus 5000 lipid molecules in total. At the initial state, the anionic and neutral lipids are homogeneously mixed in the vesicle. We first calculated the dynamics to form a phase-separated domain, subsequently reversed the direction of the DC electric field, and additionally calculated to observe the domain response. We adopted sufficiently long durations both for the domain formation $t_\mathrm{df}=10.0\times7500\tau$ and for the domain response after the reversal of the DC electric field $t=10.0\times7500\tau$. In the following, the time is represented with the unit of $7500\tau$ for simplicity. Calculations were performed five times for each strength of the DC electric field to ensure reproducibility.

\section{RESULTS}
\label{results}

First, we checked the effect of the DC electric field on the domain formation. Figure~\ref{fig1}(a) shows the time course of the domain formation of anionic lipids with no electric field, i.e., $E=0.0$, from $t_\mathrm{df}=0.0$ to $t_\mathrm{df}=10.0$ as the reference behavior. The anionic lipids rapidly assembled into a domain, and the position of the domain fluctuated with time. Figure~\ref{fig1}(b) shows the time course of the domain formation under a DC electric field $E=-0.01$. Since the anionic lipids experienced the potential force in the opposite direction of that of the electric field, they assembled toward the ``top'' of the vesicle (positive $z$-direction). Once the domain formed at the top, the domain position was fixed. The potential force further pulled the domain, and the vesicle slightly elongated and moved along the $z$-direction.

\begin{figure}[!b]
    \centering
    \includegraphics{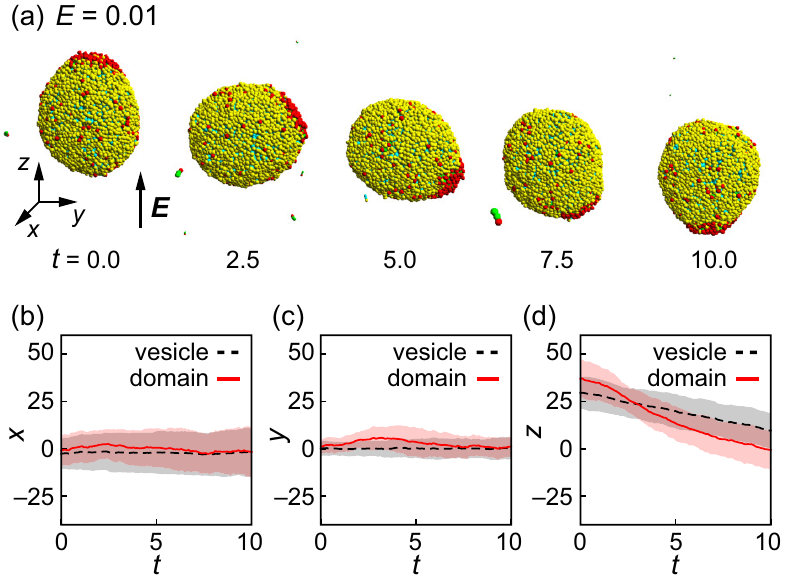}
    \caption{\label{fig2}(a) Typical snapshots of the domain dynamics after the reversal of the direction of the electric field $\mathbf{E}$ with the strength $E = 0.01$. The vesicle at $t_\mathrm{df} = 10.0$, shown in Fig.~\ref{fig1}, is used as the initial configuration, and this moment is newly set as $t = 0.0$. Calculation time $t$ is represented with the unit of $7500\tau$. (b--d) Position of the CM of the vesicle (black) and that of the anionic lipid domain (red) in Cartesian coordinates $(x,y,z)$. Shaded regions represent the corresponding standard deviations in five trials.}
\end{figure}

Afterward, we reversed the direction of the electric field. Figure~\ref{fig2}(a) shows the time course of the domain dynamics after the reversal of the direction of the electric field. Here, we set the initial configuration at $t=0.0$ to the vesicle at $t_\mathrm{df}=10.0$, shown in Fig.~\ref{fig1}, and reversed the sign of the electric field from $E=-0.01$ to $E=0.01$. The anionic lipid domain started to move along the meridian of the vesicle and finally reached the bottom of the vesicle in $t=10.0$. Figures~\ref{fig2}(b)--(d) show the position of the center of mass (CM) of the vesicle and that of the anionic lipid domain in Cartesian coordinates $(x,y,z)$. While both the vesicle and the domain fluctuated but hardly moved in the $(x,y)$ plane, they moved along the $z$-direction; The vesicle and the domain moved toward the opposite direction of the electric field. Since the domain orientation was reversed during the motion, the domain position overtook the vesicle CM at around $t\simeq3.0$, as shown in Fig.~\ref{fig2}(d).

\begin{figure}[!b]
    \centering
    \includegraphics{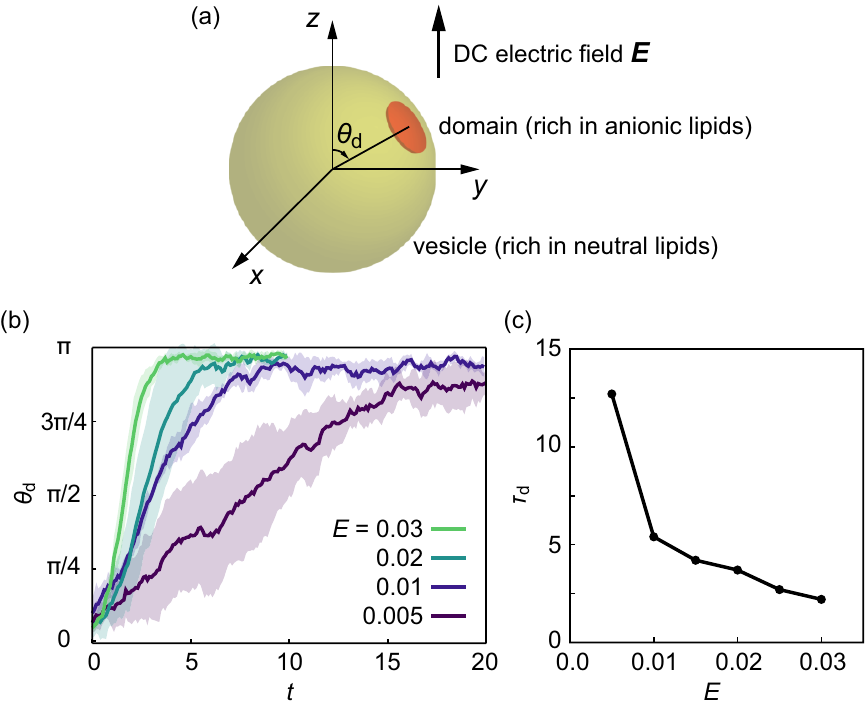}
    \caption{\label{fig3}(a) Definition of the domain orientation $\theta_\mathrm{d}$. (b) Domain orientation $\theta_\mathrm{d}$ after the reversal ($t = 0$) of the direction of the electric field $\mathbf{E}$ with various strengths $E = 0.005$, $0.01$, $0.02$, and $0.03$. Calculation time $t$ is represented with the unit of $7500\tau$. Shaded regions represent the corresponding standard deviations in five trials. (c) Characteristic time $\tau_\mathrm{d}$ of the domain motion for various $E$.}
\end{figure}

Figure~\ref{fig3} shows the dependence of the domain motion on the strengths of the reversed electric field $\mathbf{E}$. Here, we set the Cartesian coordinates $(x,y,z)$ in which the origin coincides with the vesicle CM and measured the orientation of the anionic lipid domain by the polar angle $\theta_\mathrm{d}$, as illustrated in Fig.~\ref{fig3}(a). Figure~\ref{fig3}(b) shows the time developments of the orientation of the domain for $E=0.005, 0.01, 0.02,$ and $0.03$. At $t=0$, the polar angle $\theta_\mathrm{d}$ has a small but finite value due to the thermal fluctuations of the domain. For the same reason, at sufficiently large $t$, $\theta_\mathrm{d}$ converged to a value slightly smaller than $\pi$. The increasing rate of $\theta_\mathrm{d}$ depends on the strength of the electric field $E$; the larger $E$ results in the higher increasing rate of $\theta_\mathrm{d}$.Figure~\ref{fig3}(c) shows the characteristic times $\tau_\mathrm{d}$ for the increase in $\theta_\mathrm{d}$, which was defined as the time $t(\theta_\mathrm{d}=3\pi/4)$. The characteristic time $\tau_\mathrm{d}$ significantly varies by an order of magnitude depending on the field strength $E=0.005$--$0.03$. Note that the position of the anionic lipid domain randomly fluctuated under the weaker electric fields $E\leq0.001$, and the domain was pulled out and separated from the vesicle due to the strong potential force under the stronger electric fields $E\geq0.035$.

\begin{figure}[!b]
    \centering
    \includegraphics{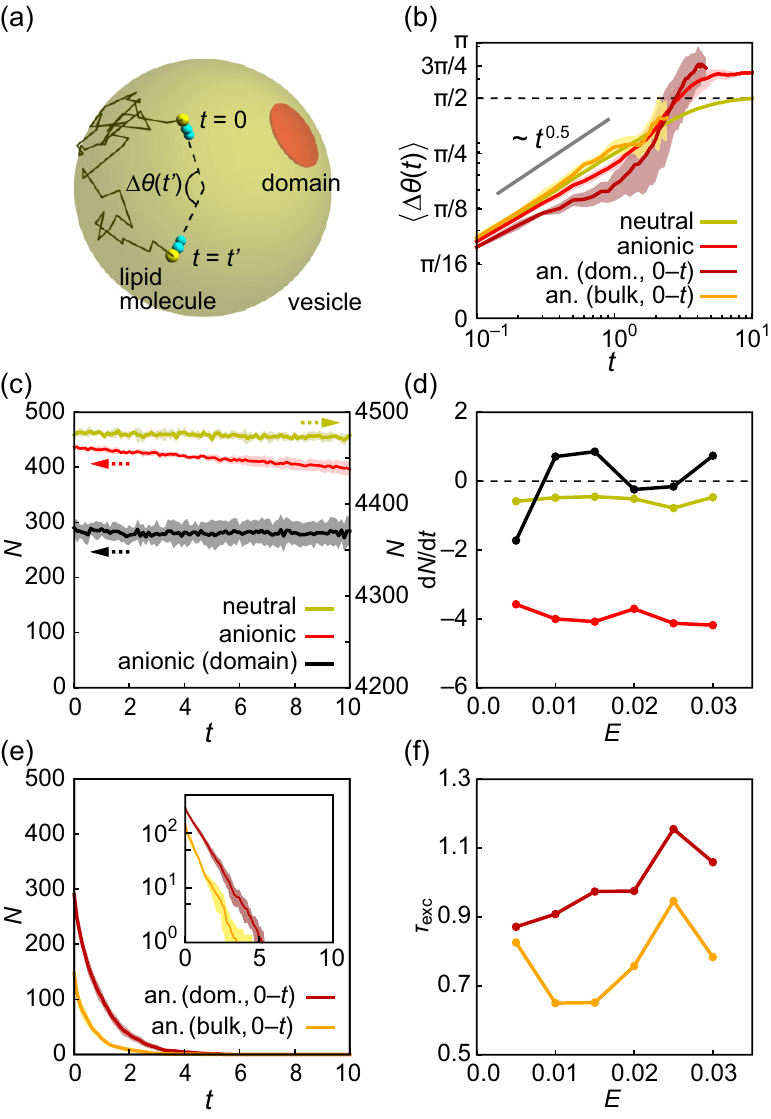}
    \caption{\label{fig4}(a) Schematic of the angle deviation $\Delta\theta(t)$ of a lipid molecule measured from the reversal of $\mathbf{E}$ at $t = 0$. (b) Mean angle deviation $\left<\Delta\theta(t)\right>$ of lipids for $E=0.02$. Neutral lipids (yellow), anionic lipids (red), and anionic lipids which continued staying in the domain (dark red) and bulk (orange) for the time duration $0$--$t$. Shaded regions represent the corresponding standard deviations in terms of lipid molecules. Calculation time $t$ is represented with the unit of $7500\tau$. (c) The number of lipids $N$ for $E=0.02$. Neutral lipids in the vesicle (yellow), anionic lipids in the vesicle (red), and anionic lipids in the domain (black). Dashed arrows indicate the corresponding vertical axes. (d) Decreasing rate of the numbers of lipids $\mathrm{d}N/\mathrm{d}t$ for various $E$. Colors are the same as (c). (e) The numbers of the anionic lipids which stayed in the domain (dark red) and bulk (orange) for the time duration $0$--$t$. Inset shows the logarithmic plot. (f) Exchange time $\tau_\mathrm{exc}$ for various $E$.}
\end{figure}

To check the detailed molecular dynamics during the motion of the domain under the reversed DC electric field, we evaluated the motility of individual lipid molecules by the deviation within the membrane. Figure~\ref{fig4}(a) shows the schematic of the angle deviation $\Delta\theta(t)$ of a lipid molecule, defined as the difference in the orientations of a lipid molecule observed from the vesicle CM at $t=0$ and $t=t^\prime$. We calculated the mean angle deviation $\left<\Delta\theta(t)\right>$ for all neutral lipids in the vesicle, all anionic lipids in the vesicle, and anionic lipids which stayed in the domain or bulk for the time duration $0$--$t$ as shown in Fig.~\ref{fig4}(b), where $\left<\cdot\right>$ denotes the average over each lipid species in five vesicles. For neutral lipids, the mean angle deviation increased from $0$ to $\pi/2$ during the motion of the anionic lipid domain. The double logarithmic plot shows the mean angle deviation $\left<\Delta\theta(t)\right>$ converges to $\pi/2$ through the power law $\sim t^{0.5}$. Considering that the random walk results in the mean square displacement $\left<\{R\Delta\theta(t)\}^2\right>\propto t$, where $R$ is the vesicle radius, and that the mean angle deviation in a random walk over the vesicle converges to $\pi/2$, this result indicates that the neutral lipids randomly moved over the vesicle. The mean angle deviation of the anionic lipids increased from $0$ to $\approx5\pi/8$, which is larger than $\pi/2$, indicating the net directional transport of the anionic lipids. The curve $\left<\Delta\theta(t)\right>$ roughly obeys the power law $\sim t^{0.5}$ within a short time $t\sim1\times10^{-1}$, deviates below the power law around an intermediate time $t\sim1\times10^0$, and distinct increases after a long time $t>2\times10^0$. To decipher the contribution of the molecular fluctuations and the unidirectional domain motion, we plotted the mean angle deviation $\left<\Delta\theta(t)\right>$ for the anionic lipids which continued staying in the domain or in the surrounding bulk for the time duration $0$--$t$. While $\left<\Delta\theta(t)\right>$ for the anionic lipids in the bulk follows that for the random motion seen in the neutral lipids, $\left<\Delta\theta(t)\right>$ for the anionic lipids in the domain clearly shows a plateau around $t\sim1\times10^0$ followed by drastic increase similar with the increase in the domain orientation $\theta_\mathrm{d}$ (Fig.~\ref{fig3}(b)). If the direction of the electric field is kept and thus the domain is continued trapped at the top of the vesicle, the drastic increase after a long time is not observed. The increase with the power law $\sim t^{0.5}$, plateau, and drastic increase are thus attributable to the diffusion within a domain, confinement of a domain boundary, and domain transportation, respectively.

To confirm further details of the dynamics of the lipid molecules, we counted the total numbers $N$ for neutral and anionic lipids in the vesicle and that for the anionic lipids in the domain (Fig.~\ref{fig4}(c)). The number of the neutral lipids was almost constant. The total number of anionic lipids in the vesicle slightly decreased through gradual dropouts from the vesicle due to the relatively shorter attractive interaction with $w_\mathrm{c}/\sigma=1.5$ for the neutral-anionic lipid pairs compared to that with $w_\mathrm{c}/\sigma=1.7$ for the neutral-neutral and anionic-anionic pairs. On the other hand, the number of domain-forming anionic lipids, and thus the domain size, remained almost constant. We also found such tendencies in the decreasing rate $\mathrm{d}N/\mathrm{d}t$ for various $E$, as shown in Fig.~\ref{fig4}(d). The decreasing rate for the total anionic lipids was negative, and the rates for the domain-forming anionic lipids and neutral lipids were almost $0$, i.e., the numbers of these lipids were almost constants. Interestingly, the domain-forming anionic lipids were not confined in the domain during the domain motion. We also counted the number of anionic lipids that continued staying in the domain or in the bulk in the time duration $0$--$t$, as shown in Fig.~\ref{fig4}(e). The linear decrease in the logarithmic plot indicates that the numbers of these lipids exponentially decreased to be exchanged between the domain and the bulk. Figure~\ref{fig4}(f) shows the characteristic exchange time $\tau_\mathrm{exc}$ as the decreasing time of the exponential function $N(t)=N(0)\exp(-t/\tau_\mathrm{exc})$, which were on the same order of $\tau_\mathrm{exc}\sim1$ for various $E$. Considering that the characteristic timescale of the domain motion $\tau_\mathrm{d}$ significantly varied depending on the field strength $E$ (Fig.~\ref{fig3}(c)), the exchange time $\tau_\mathrm{exc}$ of the lipid molecules has no significant correlation with the field strength $E$.

\section{DISCUSSION}
\label{discussion}

The anionic lipid domain unidirectionally moves to the opposite direction of the DC electric field in the surrounding bulk rich in the disordered neutral lipids. During the domain motion, the domain keeps its size but exchanges a considerable amount of the domain-forming anionic lipids across the domain boundary. While the exchange of the individual anionic lipid molecules between the domain and the bulk is almost independent of $E$ under the thermal fluctuations (Fig.~\ref{fig4}(f)), the collective motion as the domain dynamics strongly depends on the electric field strength $E$ (Fig.~\ref{fig3}(b)). Our coarse-grained MD simulation revealed the hierarchical behaviors in response to the applied electric field.

\begin{figure}[!t]
    \centering
    \includegraphics{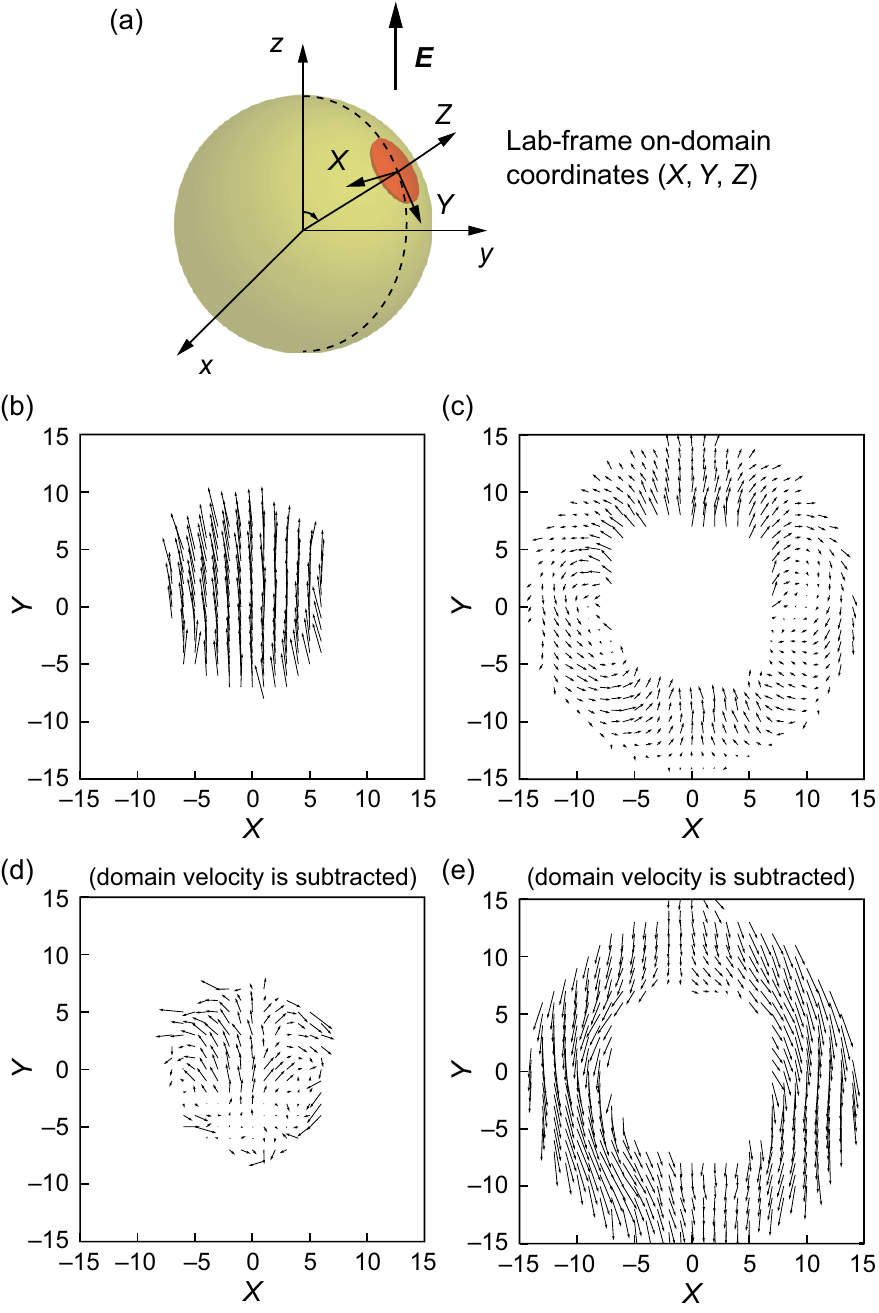}
    \caption{\label{fig5}(a) Schematic of Lab-frame on-domain coordinates $(X, Y, Z)$. Lab-frame velocity fields (b) in the domain and (c) in the surrounding bulk plotted in the projected plane $(X, Y)$ for $E=0.03$. Corresponding velocity fields (d) in the domain and (e) in the surrounding bulk where the domain velocity is subtracted.}
\end{figure}

To capture more details about the collective domain motion, we analyze the mean velocity field of the lipids inside and around the domain. As shown in Fig.~\ref{fig5}(a), we set the lab-frame on-domain right-handed Cartesian coordinates $(X,Y,Z)$ in which its origin coincides with the instantaneous domain CM and the direction of the $Y$-axis coincides the direction of the domain motion along the corresponding meridian. The basis vectors $\mathbf{e}_Y$ and $\mathbf{e}_Z$ are chosen as $\mathbf{e}_\theta$ and $\mathbf{e}_r$ in the standard spherical coordinates, respectively, and $\mathbf{e}_X = \mathbf{e}_Y\times\mathbf{e}_Z$. Figures~\ref{fig5}(b) and \ref{fig5}(c) show the mean vector fields observed in the lab-frame coordinates $(X, Y)$ for $E=0.03$. Here, the lipid velocity vectors in the 3D space $(X, Y, Z)$ are projected onto the 2D $(X, Y)$ plane for visualization. To extract the contribution of collective lipid motions from the data under the fluctuations, the velocity fields are averaged over five trials. In addition, the mean velocity field in a stationary state in terms of domain motions is subtracted as background lipid fluctuations from the mean velocity field in a moving state of a domain. 
In the lab-frame projected plane $(X, Y)$, the domain unidirectionally moves toward the positive $Y$-direction (Fig.~\ref{fig5}(b)), and the surrounding bulk exhibits a convection in a form of a 2D source dipole (Fig.~\ref{fig5}(c)), in which the source and sink locate at the front and back of the moving domain, respectively. Figures~\ref{fig5}(d) and \ref{fig5}(e) show the corresponding velocity fields in the domain and in the surrounding bulk where the velocity of the domain CM is subtracted, respectively. If we subtract the velocity of the domain CM, i.e., if we measure the lipid velocities from the frame moving at the velocity of the domain CM, the velocity field inside the domain exhibits a pair of convective rolls as shown in Fig.~\ref{fig5}(d). The velocity field in the surrounding bulk where the velocity of the domain CM is subtracted exhibits the fields similar with a flow past a fixed circular obstacle in this moving frame. Such a characteristic velocity field associated with the domain motion under a DC electric field has not been identified in the experiment\cite{Zendejas2011}. Our coarse-grained MD simulation suggests that the hydrodynamic nature of the lipid membrane plays an important role in determining the details in the dynamic response of the self-assembled domain to a DC electric field.

The reversal of the direction of the DC electric field induces the motion of the anionic lipid domain toward the opposite direction of the electric field. The anionic lipid domain moves in the surrounding neutral lipids, leading to the formation of a characteristic 2D source dipole of the mean velocity field around the domain. The 2D source dipole is reminiscent of a hydrodynamic source dipole, which appears in 2D hydrodynamic systems at a low Reynolds number, such as a disk-shaped droplet dragged in a quasi-2D channel flow\cite{Beatus2012}. To construct a theoretical framework for orienting the domain by a DC electric field, therefore, we considered a quasi-2D hydrodynamic drag problem. 

The motion of an anionic lipid domain is driven by a DC electric field. The potential force exerted on a circular domain under the DC electric field can be described with the instantaneous direction $\mathbf{e}_Y$ as
\begin{equation}
    \mathbf{F}_\mathrm{EF} = \pi a^2\Sigma E\sin\theta_\mathrm{d}\mathbf{e}_Y,
\end{equation}
where $a$ and $\Sigma$ are the radius and the surface charge density of the domain. To describe the quasi-2D mobility of the domain in the membrane, we assumed a hydrodynamic system consisting of a 2D planar incompressible liquid membrane sandwiched by a 3D solvent\cite{Saffman1975,Ramachandran2011}. In our coarse-grained MD simulation, the viscosity of the 3D solvent $\eta_\mathrm{s}$ and the viscosity of the 2D membrane $\eta_\mathrm{m}$ correspond to the viscosity coefficient $\eta$ which appears in the fluctuation-dissipation theorem in Eq.~(\ref{eq:fdt}) and effective viscosity originating from the interactions between lipid molecules, respectively. In such a condition, the mobility of a circular fluid domain embedded in the 2D membrane, which is called Saffman-Delbr\"{u}ck mobility, is described for $\eta_\mathrm{s}\ll\eta_\mathrm{m}$ as
\begin{equation}
    b_\mathrm{T} = \frac{1}{4\pi\eta_\mathrm{m}h}\left(\ln\frac{\eta_\mathrm{m}h}{\eta_\mathrm{s}a}-\gamma+\frac{1}{2}\right),
\end{equation}
where $h$ is the thickness of the membrane and $\gamma\approx0.5772$ is the Euler's constant\cite{Saffman1975,Saffman1976,Cicuta2007}. Considering the friction coefficient $\lambda_\mathrm{T}=b_\mathrm{T}^{-1}$ against the slow domain motion, the balance equation becomes
\begin{equation}
    \mathbf{F}_\mathrm{EF} - \lambda_\mathrm{T}R\frac{\mathrm{d}\theta_\mathrm{d}}{\mathrm{d}\hat{t}}\mathbf{e}_Y = \mathbf{0},
\end{equation}
where $\hat{t}$ and $R$ are the theoretical time and the radius of the vesicle, respectively. Integrating this equation reads
\begin{gather}
    f(\theta_\mathrm{d})\equiv\ln\frac{\cot\frac{\theta_\mathrm{d}}{2}}{\cot\frac{\theta_0}{2}} = -\frac{\hat{t}}{\hat{\tau}},\label{eq:theta_d_0} \\
    \hat{\tau}^{-1} = \frac{\Sigma Ea^2}{4\eta_\mathrm{m}hR}\left(\ln\frac{\eta_\mathrm{m}h}{\eta_\mathrm{s}a}-\gamma+\frac{1}{2}\right),\label{eq:theta_d_1}
\end{gather}
or equivalently,
\begin{equation}
    \theta_\mathrm{d}(\hat{t}) = 2\cot^{-1}\left[\cot\frac{\theta_0}{2}\exp\left(-\frac{\hat{t}}{\hat{\tau}}\right)\right],\label{eq:theta_d_2}
\end{equation}
where $\theta_\mathrm{d}(0) = \theta_0$ is the initial polar angle of the domain orientation. 
\begin{figure}[!t]
    \centering
    \includegraphics{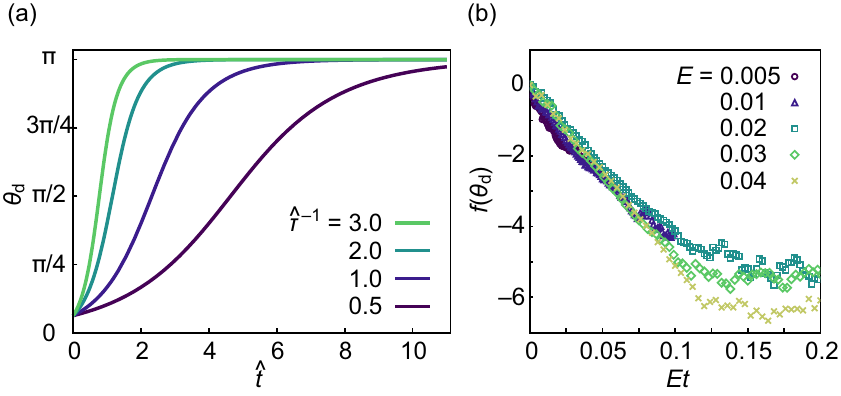}
    \caption{\label{fig6}(a) Theoretical curves of the domain orientation $\theta_\mathrm{d}$ after the reversal ($\hat{t} = 0$) of the direction of the electric field $\mathbf{E}$ for various strength parameters $\hat{\tau}^{-1} = 0.5$, $1.0$, $2.0$, and $3.0$. (b) $f(\theta_{\mathrm{d}})$ versus $Et$ for the data from coarse-grained MD simulation.}
\end{figure}
Figure~\ref{fig6}(a) shows the theoretical curves of the domain orientation $\theta_\mathrm{d}(\hat{t})$ described in Eq.~(\ref{eq:theta_d_2}) after the reversal of the direction of the electric field at $\hat{t}=0$ for various strengths of the electric field. Here, we plotted for $\hat{\tau}^{-1}=0.5$, $1.0$, $2.0$, and $3.0$, which are proportional to the strength of the electric field. For the initial orientation, we set $\theta_0=\pi/16$ by considering the fluctuation of the domain orientation. As we can see in Fig.~\ref{fig6}(a), the theoretical curves qualitatively reproduced the results of coarse-grained MD simulation shown in Fig.~\ref{fig3}(b) in both the time development of the domain orientation $\theta_\mathrm{d}(t)$ and its dependence on the strength of the electric field $E$. For further validation, we also plotted $f(\theta_{\mathrm{d}})$, defined in Eq.~(\ref{eq:theta_d_0}), obtained from the coarse-grained MD simulation, as shown in Fig.~\ref{fig6}(b). According to Eqs.~(\ref{eq:theta_d_0}) and (\ref{eq:theta_d_1}), the data for various strengths $E$ should collapse on a linear line as a function of $Et$. Figure~\ref{fig6}(b) shows that the data for $Et<0.1$ clearly collapse on a linearly decreasing line irrespective of the strength $E$. For $Et>0.1$, which corresponds to $\theta_\mathrm{d}\approx0.9$, the domain almost reaches the opposite pole, and the data deviate from the collapsed line with thermal fluctuations. The finally stabilized position $\theta_\mathrm{d}(Et\simeq0.2)$ shows a weak positive correlation with the strength $E$, resulting from competition with the thermal fluctuations. Although the microscopic dynamics is complex, as seen in Fig.~\ref{fig4}, the mesoscopic domain-scale dynamics can be predicted through the continuum description for the quasi-2D fluidic membrane.

Generally, domain motion can be affected by the contributions of both the electrostatic potential force exerted on the charged lipids and the frictional force due to the electroosmotic flow induced by the bulk ions accumulated near the membrane\cite{Stelzle1992}. In the previous experimental demonstration\cite{Zendejas2011}, phase-separated domains rich in negatively charged lipids oriented toward the positive electrode, and the positively charged domains oriented toward the negative electrode. This experiment suggests that the electrostatic force, rather than the electroosmotic flow, is dominant for orienting the domain in the lipid bilayer membrane. In the present coarse-grained MD simulation, the effects of bulk ions are implicitly included in the screening effect on the electrostatic interaction Eq.~(\ref{eq:V_rep}) between the anionic head groups, and thus potential modifications to the present results by the effects of the bulk electroosmotic flow are neglected. For more precise quantitative predictions of the domain dynamics, simulations with explicit ions are needed in future work.

\section{CONCLUSION}
\label{conclusion}
Using coarse-grained MD simulations, we studied the dynamical lateral transport of a phase-separated domain rich in anionic lipids in a lipid bilayer vesicle under an externally applied DC electric field. Under the potential force of the electric field, the anionic lipid domain is trapped in the opposite direction of that of the electric field, and the domain was transported along the meridian of the vesicle after the reversal of the electric field. The domain dynamics significantly correlates on the strength of the electric field. During the domain motion, the domain-forming anionic lipids were rapidly exchanged with those in the surrounding bulk, while the domain size remained almost constant. The exchange rate of individual anionic lipids between the domain and the bulk is almost independent of the electric field strength. The mean velocity field in the surrounding bulk exhibited a 2D source dipole with the source and sink at the front and back of the moving domain, respectively, indicating that the 2D hydrodynamic nature determines the domain dynamics as the collective motion of the domain-forming anionic lipid, even with the strong fluctuations. Based on the results obtained in the coarse-grained MD simulation, we described the domain dynamics by the balance equation between the potential force and hydrodynamic frictional force, which well explained the time development of the domain position as well as its field-strength dependence. The present findings not only demonstrate orienting the functional domains by external fields in a predictable way but also contribute to a fundamental understanding of the lateral transport phenomena of hierarchical structures in 2D interfaces, especially in biomembranes.

\begin{acknowledgments}
Calculations were performed using the parallel computer ``SGI UV3000'' at the Research Center for Advanced Computing Infrastructure at JAIST and Supercomputer Center at the Institute for Solid State Physics in the University of Tokyo. The research was supported by JSPS KAKENHI Grant Numbers JP21K13891 (H.I.) and JP19H05718 (Y.H.), and JSPS and MESS Japan-Slovenia Research Cooperative Program Grant Number JPJSBP120215001.
\end{acknowledgments}

\bibliography{reference_HI.bib}

\end{document}